\documentclass[twocolumn]{aastex631}
\usepackage{amsmath}
\usepackage{graphicx}
\usepackage{natbib}
\usepackage{xcolor}
\usepackage{ulem}
\usepackage{float}
\graphicspath{ {./figures/} }

\graphicspath{{./}{figures/}}

\begin{document}

\title{A Size Estimate for Galaxy GN-z11}

\author[0000-0002-3498-9086]{James O. Baldwin}
\affiliation{Department of Astrophysical and Planetary Sciences, University of Colorado Boulder, Boulder, CO 80309, USA}

\author[0000-0002-7524-374X]{Erica Nelson}
\affiliation{Department of Astrophysical and Planetary Sciences, University of Colorado Boulder, Boulder, CO 80309, USA}

\author[0000-0002-9280-7594]{Benjamin D. Johnson}
\affiliation{Center for Astrophysics | Harvard \& Smithsonian, Cambridge MA 02138 USA}

\author[0000-0001-5851-6649]{Pascal A. Oesch}
\affiliation{Departement d’Astronomie, Universitè de Genève, 51 Ch. des Maillettes, CH-1290 Versoix, Switzerland}
\affiliation{Cosmic Dawn Center (DAWN), Niels Bohr Institute, University of Copenhagen, Jagtvej 128, KØbenhavn N, DK-2200, Denmark}

\author[0000-0002-8224-4505]{Sandro Tacchella}
\affiliation{Kavli
Institute for Cosmology, University of Cambridge, Madingley Road, Cambridge, CB3 0HA, UK}
\affiliation{Cavendish Laboratory, University of Cambridge, 19 JJ Thomson Avenue, Cambridge, CB3 0HE, UK}

\author[0000-0002-8096-2837]{Garth D. Illingworth}
\affiliation{UCO/Lick Observatory, University of California, Santa Cruz, 1156 High Street, Santa Cruz, CA 95064, USA}

\author{Justus Gibson}
\affiliation{Department of Astrophysical and Planetary Sciences, University of Colorado Boulder, Boulder, CO 80309, USA}

\author{Abby Hartley}
\affiliation{Department of Astrophysical and Planetary Sciences, University of Colorado Boulder, Boulder, CO 80309, USA}eric

\begin{abstract}

GN-z11 is the highest redshift galaxy spectroscopically confirmed with the Hubble Space Telescope (HST). Previous measurements of the effective radius of GN-z11 utilized \textsc{galfit}, which is not optimized to measure structural parameters for such a faint, distant object. Using a new software program called \textsc{forcepho} on HST data for the first time, we derive a size from images in the F160W band obtained both from the complete CANDELS survey and additional midcycle observations in order to contribute to the knowledge base on the size evolution, size-luminosity, and size-mass relation of early galaxies. We find a half-light radius mean of 0”.036 ± 0”.006 corresponding to a physical size of 0.15 ± 0.025 kpc. This size, smaller than the point spread function, is dramatically smaller than previous estimates with shallower HST data using galfit but consistent with recent measurements using forcepho on new JWST data \citep{https://doi.org/10.48550/arxiv.2302.07234}. Such a small size, combined with the JWST/NIRSpec spectroscopic observations \citep{maiolino2023small}, suggests that GN-z11's high luminosity is  dominated by an AGN. 

\end{abstract}

\section{Introduction} \label{sec:intro}

High-redshift galaxies represent some of the first structures to form in our universe. By understanding the physical properties of these galaxies, we can refine our theoretical models and better interpret our observations \citep{doi:10.1146/annurev-astro-081915-023417, Dayal_2019}. These galaxies, particularly those at $z \sim 9-11$, also existed at a time when the universe experienced a shift from a neutral to ionized state \citep[e.g.,][]{Oesch_2016}. Constraints on their physical properties yield insights into their role in reionizing the universe \citep[e.g.,][]{Bunker_2010, Robertson_2013}.

First identified as GNS-JD2 in the HST/NICMOS data, GN-z11 was initially considered a possible dropout candidate due to its detection being limited to the $1.6 \hspace{1mm}\mu m$ wavelength \citep{Bouwens_2010}. Its proximity to another source made further conclusions difficult until 2014 when \cite{Oesch_2014} classified it (under the designation GN-z10-1) as a $z\sim9-10$ galaxy, and later determined its redshift, through grism spectroscopy, to be $z = 11.09_{-0.12}^{+0.08}$ \citep{Oesch_2016}, a measurement more accurately constrained with more recent NIRSpec data to be $z=10.603$ \citep{bunker2023jades}. One of the most distant galaxy observed with HST, GN-z11 is located in the CANDELS DEEP area in GOODS-North at (RA,DEC) = (12:36:25.46, +62:14:31.4).

\cite{Holwerda_2015} measured the structure of a sample of $z\sim 9-10$ galaxies, including GN-z11 (designated GN-z10-1 in their paper), using the 2D fitting algorithm \textsc{galfit} \citep{Peng_2002, Peng_2010}. The two main drawbacks of \textsc{galfit} when determining the physical properties of galaxies at the distance and luminosity of GN-z11 are (a) the lack of posterior distributions complicate the interpretation of the uncertainty and (b) the software is not optimized for faint, distant objects. \cite{Holwerda_2015} report a size for GN-z11 of $0.6 \pm 0.3$kpc.
The uncertainty reported is statistical, however, and based on an estimate of models with similar estimated structural properties (see \cite{Ono_2013} for a description of \textsc{galfit} uncertainties and biases). Further, their measurement was obtained by fixing the S\'ersic index, $n$, and the axis-ratio, $q$, to limit the statistical uncertainty. These fixed parameter models have allowed for initial estimations of the size of GN-z11, but may suffer from systematic uncertainties from modeling choices. Here, we build on previous work with \textsc{galfit} using a new Bayesian size-fitting software, \textsc{forcepho}, which is optimized for faint, blended objects and provides robust uncertainties (Johnson et al., in prep).

In this research note, we present a size measurement of GN-z11 using \textsc{forcepho}. In Section~\ref{sec:style}, we present our observations and the methodology for carrying out the measurement, with a description of the software package \textsc{forcepho}. We present our size-measurement in Section~\ref{sec:results} and the significance of this measurement, along with a discussion of more recent findings, is presented in Section~\ref{sec:results}. In Section~\ref{sec:conclusions} we provide a brief conclusion.


\begin{figure*}
    \centering
    \includegraphics[width=0.7\textwidth]{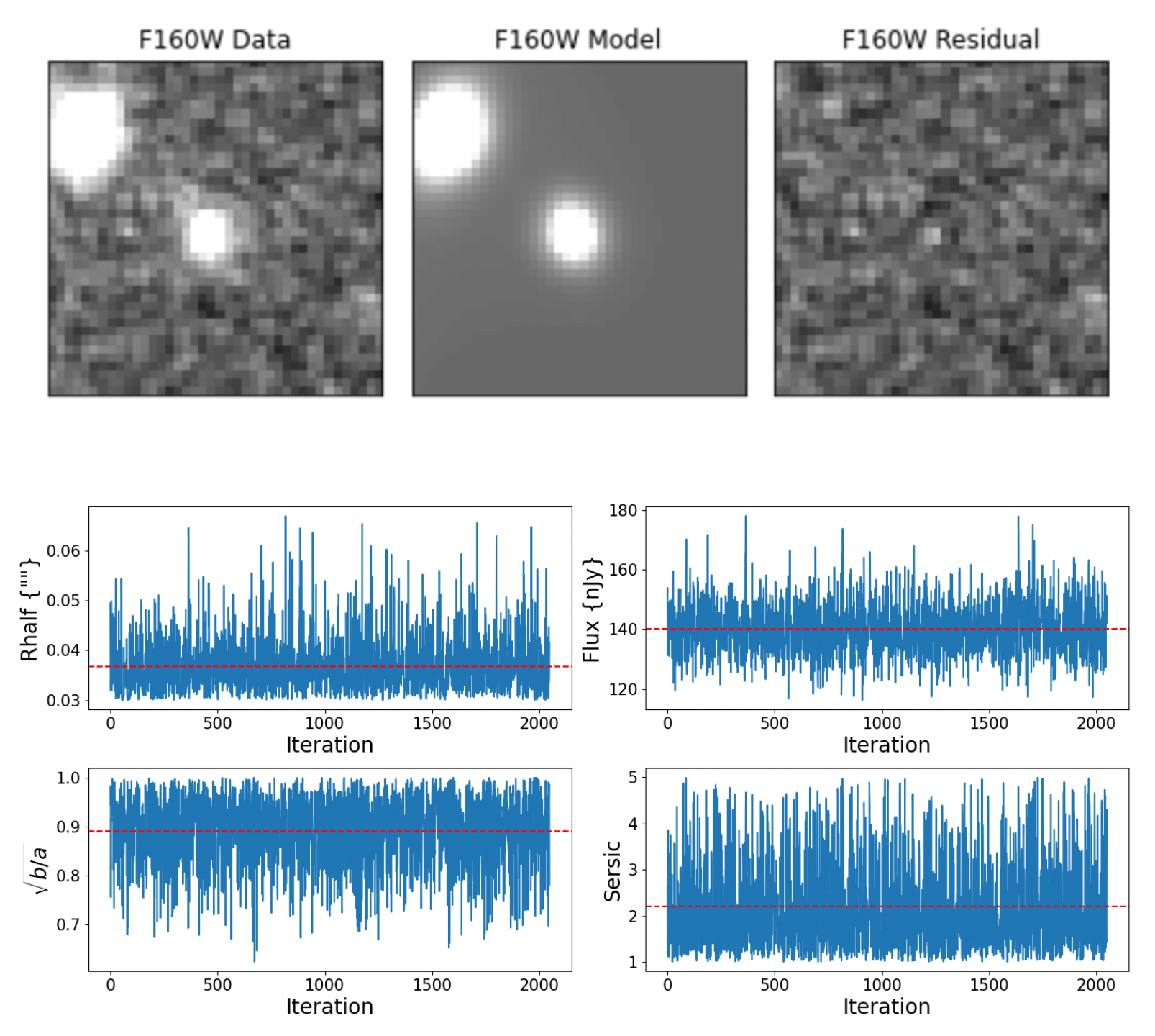}
    \caption{TOP: Results of the \textsc{forcepho} modeling after optimization and sampling. The far left shows the original HST data image, the middle image shows the model constructed from the PSF, and the final image is the residual constructed from subtracting the model from the data. BOTTOM: Posteriors generated by \textsc{forcepho} after optimization and sampling. The MCMC sampling was performed over 2000 iterations, with the dotted red line representing the mean value for each posterior.}
    \label{fig:outputs}
\end{figure*}



\section{Data Analysis} \label{sec:style}

GN-z11 was imaged using the WFC3/IR camera aboard HST. Existing imaging of the GOODS-N field from the CANDELS survey consisted of $\sim5$ orbits covering the central $\sim65$ arcmin$^2$ of the GOODS-N field \citep{Oesch_2014}. 7 additional orbits in F160W were obtained as a result of a mid-cycle proposal for a total of 12 orbits, reaching 27.7 mag ($5\sigma$) (PI:Oesch, PID: 15977). See \cite{Oesch_2014} and \cite{Illingworth_2013} for a description of the data reduction.  



We fit the structural parameters of GN-z11 using \textsc{forcepho}, which infers the fluxes and morphological parameters of galaxies via a Bayesian analysis (Johnson et al., in prep). Models are generated for our sources which are compared to the observations via a likelihood function as the posterior is sampled. Crucially, \textsc{forcepho} is differentiable, meaning that gradients of the likelihood function can be taken, allowing for significant optimization in the posterior sampling. Additionally, \textsc{forcepho} utilizes Gaussians to approximate the point spread functions (PSFs) --- constructed empirically from a stack of point sources in the HLF imaging --- and the S\'ersic profiles. The sums of these Gaussian parameters are used for convolution, which greatly increases efficiency. We use \textsc{forcepho} to measure the size and other structural parameters of GN-z11 in the F160W band. In addition to the half-light radius, we also infer S\'ersic index and the axis-ratio, not setting them to fixed values as in \cite{Holwerda_2015}. The major source of uncertainty in our measurements come from the similarity of the galaxy size to the size of the PSF. 

\section{Results \& Discussion} \label{sec:results}
We measure a half-light radius of $0".036 \pm 0".006$ which corresponds to a physical size of $0.15\pm 0.025$ kpc, assuming a redshift of $z = 10.603$ as measured by \cite{bunker2023jades}. In Fig.~\ref{fig:outputs} (top), we show the image, model, and residual (data - model) showing that the \textsc{forcepho} model is a good match to the data. We infer a flux in F160W of $139.16 \pm 8.98$ nJy, less than the initial \cite{Oesch_2016} finding of $152 \pm 10$ nJy, but within the range of uncertainty. These measurements differ significantly from those obtained by \cite{Holwerda_2015} ($0.6\pm0.3$ kpc), finding of a much smaller half-light radius suggestive of a point source. The mean S\'ersic index was measured to be $2.42 \pm 1.12$, and the axis-ratio was measured to have a mean of $0.91 \pm 0.07$. These measurements are higher than those found by more recent analysis which finds ($n=0.9\pm0.1$, $q=0.67\pm0.05$) using \textsc{forcepho} on 0.9-4.4\micron\ JWST/NIRCam imaging \citep{https://doi.org/10.48550/arxiv.2302.07234}. This could be due to the increased spatial resolution and signal-to-noise ratio of the JWST obervations or an intrinsically different shape in the longer wavelength bands. Fig.~\ref{fig:outputs} (bottom) shows the posteriors from the MCMC sampling, including the mean values for each of these values.

Our measured half-light radius is significantly smaller than previous measurements based on HST imaging. This measurement is confirmed by recent analysis of data collected as part of the JWST Advanced Deep Extragalactic Survey (JADES). Utilizing NIRCam data in nine separate bands (F090W, F115W, F150W, F200W, F277W, F356W,
F410M, F444W, and F335M), analysis shows a half-light radius of $0.016" \pm 0".005$ \citep{https://doi.org/10.48550/arxiv.2302.07234}.

Given its high redshift, the remarkable luminosity of GN-z11 is hard to explain with stellar populations alone \citep{Mason_2015, mashian2015empirical, Trac_2015, Tacchella_2013}. In their paper analyzing JADES NIRSpec spectroscopy of GN-z11, \cite{bunker2023jades} examined the possible sources of photoionization and, based on emission line fluxes, suggested that an AGN is not a favored source of the luminosity, \textit{though can not be completely ruled out}. Both \cite{https://doi.org/10.48550/arxiv.2302.07234} and \cite{bunker2023jades} find evidence for a younger stellar population, with Tacchella et al. observing a weak Balmer/4000 \r{A} break and Bunker et al. observing no Balmer Break whatsoever.

More recent analysis, however, points to the presence of an AGN within GN-z11, accounting for the unresolved nuclear component (the AGN) and the disk-like component (the galaxy) revealed in Tacchella et al. \citep{maiolino2023small}. This analysis, based on data taken as part of the JADES survey, relies on a deeper spectrum of GN-z11, and reveals features consistent with the presence of an AGN: the detection of [NeIV] and CII lines, extremely high gas density, and a deep, blueshifted absoprtion trough of CIV \citep{maiolino2023small}. In particular, the high densities are fully within the realm of Broad Line Regions (BLR) of AGN, with widths consistent with Narrow Line Seyfert 1 (NLS1) AGN \citep{maiolino2023small}. 

Previous studies suggest that GN-z11 hosts a population of Wolf-Rayet stars that may account for its luminosity \citep{Cameron_2023, Charbonnel_2023, senchyna2023gnz11}. \cite{maiolino2023small} point out the absence of a strong NIV$\lambda$1718 line in conjunction with the observed NIV$\lambda\lambda$1483,1486 doublet, as well as the presence of the previously mentioned [NeIV] and CII lines, neither of which is associated with WR stars \citep{maiolino2023small}.

Based on the spectral line widths and the continuum luminosity, the authors estimate a BH mass of about $1.6 \times 10^6 M_\odot$ \citep{maiolino2023small}. A bolometric luminosity of $10^{45}$ erg/s is inferred, consistent with Super-Eddington accretion which the authors suggest is likely episodic \citep{maiolino2023small}. While it is possible this accretion rate extends to the past and the AGN was generated from a stellar black hole seed formed at $z\sim$12-15, more massive seed scenarios associated with direct black hole collapse (DBHC) and the merging of nuclear clusters would have provided easier routes to a BH of this mass ($1.6\times 10^6 M_\odot$) at this time ($z=10.6$) \citep{maiolino2023small}.

\section{Conclusions} \label{sec:conclusions}

In summary, we performed a new size-measurement for GN-z11, the highest redshift galaxy spectroscopically confirmed by HST, using the Bayesian fitting software \textsc{forcepho}. Our physical size of $0.15\pm 0.1$ kpc is much smaller than previous measurements with \textsc{galfit}, but consistent with measurements based on newer JWST/NIRCam data \citep{https://doi.org/10.48550/arxiv.2302.07234}. The size of GN-z11, along with its redshift and luminosity, present interesting possibilities for the physical processes at play in such an early galaxy \citep{bunker2023jades,https://doi.org/10.48550/arxiv.2302.07234}. 
Our measurement of a very compact size in GN-z11 is consistent with WST/NIRSpec observations \citep{maiolino2023small} suggesting that at least some of the remarkable luminosity of this object may be driven by an AGN. 

\section{Acknowledgments}
This work is based on observations taken by the 3D-HST Treasury Program (GO 12177 and 12328) with the NASA/ESA HST, which is operated by the Association of Universities for Research in Astronomy, Inc., under NASA contract NAS5-26555. Support from HST-GO-15977, HST-GO-13871, and the CANDELS survey is gratefully acknowledged. 

\newpage
\bibliographystyle{mnras}
\bibliography{gnz11}

\begin{thebibliography}{}
\makeatletter
\relax
\def\mn@urlcharsother{\let\do\@makeother \do\$\do\&\do\#\do\^\do\_\do\%\do\~}
\def\mn@doi{\begingroup\mn@urlcharsother \@ifnextchar [ {\mn@doi@}
  {\mn@doi@[]}}
\def\mn@doi@[#1]#2{\def\@tempa{#1}\ifx\@tempa\@empty \href
  {http://dx.doi.org/#2} {doi:#2}\else \href {http://dx.doi.org/#2} {#1}\fi
  \endgroup}
\def\mn@eprint#1#2{\mn@eprint@#1:#2::\@nil}
\def\mn@eprint@arXiv#1{\href {http://arxiv.org/abs/#1} {{\tt arXiv:#1}}}
\def\mn@eprint@dblp#1{\href {http://dblp.uni-trier.de/rec/bibtex/#1.xml}
  {dblp:#1}}
\def\mn@eprint@#1:#2:#3:#4\@nil{\def\@tempa {#1}\def\@tempb {#2}\def\@tempc
  {#3}\ifx \@tempc \@empty \let \@tempc \@tempb \let \@tempb \@tempa \fi \ifx
  \@tempb \@empty \def\@tempb {arXiv}\fi \@ifundefined
  {mn@eprint@\@tempb}{\@tempb:\@tempc}{\expandafter \expandafter \csname
  mn@eprint@\@tempb\endcsname \expandafter{\@tempc}}}

\bibitem[\protect\citeauthoryear{Bouwens et~al.,}{Bouwens
  et~al.}{2010}]{Bouwens_2010}
Bouwens R.~J.,  et~al., 2010, \mn@doi [The Astrophysical Journal]
  {10.1088/0004-637x/725/2/1587}, 725, 1587

\bibitem[\protect\citeauthoryear{Bunker et~al.,}{Bunker
  et~al.}{2010}]{Bunker_2010}
Bunker A.~J.,  et~al., 2010, \mn@doi [Monthly Notices of the Royal Astronomical
  Society] {10.1111/j.1365-2966.2010.17350.x}, 409, 855

\bibitem[\protect\citeauthoryear{Bunker et~al.,}{Bunker
  et~al.}{2023}]{bunker2023jades}
Bunker A.~J.,  et~al., 2023, JADES NIRSpec Spectroscopy of GN-z11:
  Lyman-$\alpha$ emission and possible enhanced nitrogen abundance in a
  $z=10.60$ luminous galaxy (\mn@eprint {arXiv} {2302.07256})

\bibitem[\protect\citeauthoryear{Cameron, Katz, Rey  \& Saxena}{Cameron
  et~al.}{2023}]{Cameron_2023}
Cameron A.~J.,  Katz H.,  Rey M.~P.,   Saxena A.,  2023, \mn@doi [Monthly
  Notices of the Royal Astronomical Society] {10.1093/mnras/stad1579}, 523,
  3516

\bibitem[\protect\citeauthoryear{Charbonnel, Schaerer, Prantzos, Ram{\'{\i}
  }rez-Galeano, Fragos, Kuruvanthodi, Marques-Chaves  \& Gieles}{Charbonnel
  et~al.}{2023}]{Charbonnel_2023}
Charbonnel C.,  Schaerer D.,  Prantzos N.,  Ram{\'{\i} }rez-Galeano L.,  Fragos
  T.,  Kuruvanthodi A.,  Marques-Chaves R.,   Gieles M.,  2023, \mn@doi
  [Astronomy \& Astrophysics] {10.1051/0004-6361/202346410}, 673, L7

\bibitem[\protect\citeauthoryear{Dayal}{Dayal}{2019}]{Dayal_2019}
Dayal P.,  2019, \mn@doi [Proceedings of the International Astronomical Union]
  {10.1017/s1743921320001106}, 15, 43

\bibitem[\protect\citeauthoryear{Holwerda, Bouwens, Oesch, Smit, Illingworth
  \& Labbe}{Holwerda et~al.}{2015}]{Holwerda_2015}
Holwerda B.~W.,  Bouwens R.,  Oesch P.,  Smit R.,  Illingworth G.,   Labbe I.,
  2015, \mn@doi [The Astrophysical Journal] {10.1088/0004-637x/808/1/6}, 808, 6

\bibitem[\protect\citeauthoryear{Illingworth et~al.,}{Illingworth
  et~al.}{2013}]{Illingworth_2013}
Illingworth G.~D.,  et~al., 2013, \mn@doi [The Astrophysical Journal Supplement
  Series] {10.1088/0067-0049/209/1/6}, 209, 6

\bibitem[\protect\citeauthoryear{Maiolino et~al.,}{Maiolino
  et~al.}{2023}]{maiolino2023small}
Maiolino R.,  et~al., 2023, A small and vigorous black hole in the early
  Universe (\mn@eprint {arXiv} {2305.12492})

\bibitem[\protect\citeauthoryear{Mashian, Oesch  \& Loeb}{Mashian
  et~al.}{2015}]{mashian2015empirical}
Mashian N.,  Oesch P.,   Loeb A.,  2015, An Empirical Model for the Galaxy
  Luminosity and Star-Formation Rate Function at High Redshift (\mn@eprint
  {arXiv} {1507.00999})

\bibitem[\protect\citeauthoryear{Mason, Trenti  \& Treu}{Mason
  et~al.}{2015}]{Mason_2015}
Mason C.~A.,  Trenti M.,   Treu T.,  2015, \mn@doi [The Astrophysical Journal]
  {10.1088/0004-637x/813/1/21}, 813, 21

\bibitem[\protect\citeauthoryear{Oesch et~al.,}{Oesch
  et~al.}{2014}]{Oesch_2014}
Oesch P.~A.,  et~al., 2014, \mn@doi [The Astrophysical Journal]
  {10.1088/0004-637x/786/2/108}, 786, 108

\bibitem[\protect\citeauthoryear{Oesch et~al.,}{Oesch
  et~al.}{2016}]{Oesch_2016}
Oesch P.~A.,  et~al., 2016, \mn@doi [The Astrophysical Journal]
  {10.3847/0004-637x/819/2/129}, 819, 129

\bibitem[\protect\citeauthoryear{Ono et~al.,}{Ono et~al.}{2013}]{Ono_2013}
Ono Y.,  et~al., 2013, \mn@doi [The Astrophysical Journal]
  {10.1088/0004-637x/777/2/155}, 777, 155

\bibitem[\protect\citeauthoryear{Peng, Ho, Impey  \& Rix}{Peng
  et~al.}{2002}]{Peng_2002}
Peng C.~Y.,  Ho L.~C.,  Impey C.~D.,   Rix H.-W.,  2002, \mn@doi [The
  Astronomical Journal] {10.1086/340952}, 124, 266

\bibitem[\protect\citeauthoryear{Peng, Ho, Impey  \& Rix}{Peng
  et~al.}{2010}]{Peng_2010}
Peng C.~Y.,  Ho L.~C.,  Impey C.~D.,   Rix H.-W.,  2010, \mn@doi [The
  Astronomical Journal] {10.1088/0004-6256/139/6/2097}, 139, 2097

\bibitem[\protect\citeauthoryear{Robertson et~al.,}{Robertson
  et~al.}{2013}]{Robertson_2013}
Robertson B.~E.,  et~al., 2013, \mn@doi [The Astrophysical Journal]
  {10.1088/0004-637x/768/1/71}, 768, 71

\bibitem[\protect\citeauthoryear{Senchyna, Plat, Stark  \& Rudie}{Senchyna
  et~al.}{2023}]{senchyna2023gnz11}
Senchyna P.,  Plat A.,  Stark D.~P.,   Rudie G.~C.,  2023, GN-z11 in context:
  possible signatures of globular cluster precursors at redshift 10 (\mn@eprint
  {arXiv} {2303.04179})

\bibitem[\protect\citeauthoryear{Stark}{Stark}{2016}]{doi:10.1146/annurev-astro-081915-023417}
Stark D.~P.,  2016, \mn@doi [Annual Review of Astronomy and Astrophysics]
  {10.1146/annurev-astro-081915-023417}, 54, 761

\bibitem[\protect\citeauthoryear{Tacchella, Trenti  \& Carollo}{Tacchella
  et~al.}{2013}]{Tacchella_2013}
Tacchella S.,  Trenti M.,   Carollo C.~M.,  2013, \mn@doi [The Astrophysical
  Journal] {10.1088/2041-8205/768/2/l37}, 768, L37

\bibitem[\protect\citeauthoryear{Tacchella et~al.,}{Tacchella
  et~al.}{2023}]{https://doi.org/10.48550/arxiv.2302.07234}
Tacchella S.,  et~al., 2023, JADES Imaging of GN-z11: Revealing the Morphology
  and Environment of a Luminous Galaxy 430 Myr After the Big Bang,
  \mn@doi{10.48550/ARXIV.2302.07234}, \url {https://arxiv.org/abs/2302.07234}

\bibitem[\protect\citeauthoryear{Trac, Cen  \& Mansfield}{Trac
  et~al.}{2015}]{Trac_2015}
Trac H.,  Cen R.,   Mansfield P.,  2015, \mn@doi [The Astrophysical Journal]
  {10.1088/0004-637x/813/1/54}, 813, 54

\makeatother
\end{thebibliography}

\end{document}